# Quantum magnetic imaging of iron biomineralisation in teeth of the chiton *Acanthopleura hirtosa*

Julia M. McCoey[1]*, Mirai Matsuoka[1], Robert W. de Gille[1], Liam T. Hall[1], Jeremy A. Shaw[2], Jean-Philippe Tetienne[1], David Kisailus[3], Lloyd C.L. Hollenberg[1], David A. Simpson[1]

[1]School of Physics, The University of Melbourne, Melbourne 3010, Australia

[2]Centre for Microscopy, Characterization and Analysis, University of Western Australia, Perth, WA, Australia

[3]Materials Science and Engineering Program, University of California, Riverside, CA 92521, USA

**Iron biomineralisation is critical for life. Nature capitalises on the physical attributes of iron biominerals for a variety of functional, structural and sensory applications[1–5]. Although magnetism is an integral property of iron biominerals, the role it plays in their nano-assembly remains a fundamental, unanswered question. This is well exemplified by the magnetite-bearing radula of chitons. Chitons, a class of marine mollusc, create the hardest biomineral of any animal in their abrasion-resistant, self-sharpening teeth[4]. Despite this system being subjected to a range of high resolution imaging studies, the mechanisms that drive mineral assembly remain unresolved. However, the advent of quantum imaging technology provides a new avenue to probe magnetic structures directly. Here we use quantum magnetic microscopy[6], based on nitrogen-vacancy centres in diamond, to attain the first subcellular magnetic profiling of a eukaryotic system. Using complementary magnetic imaging protocols, we spatially map the principal mineral phases (ferrihydrite and magnetite) in the developing teeth of *Acanthopleura hirtosa* with submicron resolution. The images reveal previously undiscovered long-range magnetic order, established at the onset of magnetite mineralisation. This is in contrast to electron microscopy studies that show no strong common crystallographic orientation[7]. The quantum-based magnetic profiling techniques presented in this work have broad application in biology, earth science, chemistry and materials engineering and can be applied across the range of systems for which iron is vital.**

Living organisms produce iron biominerals to serve a range of structural and functional roles: to store and transport iron[1,2], for sensing and navigational purposes[3], and to produce reinforced structural materials[4,5]. Chitons represent an ideal model system to study iron biomineralization. Anatomically, they possess eight hard dorsal plates and a soft body, with a ventrally-facing mouth (Fig. 1a). Chitons require structurally hard teeth because of their feeding habits, and thus produce the hardest known biomineralised material[4]. They are omnivorous grazers, rasping food from rocky substrates with their radula, a tongue-like organ with rows of curved teeth. The fully developed teeth possess material properties equivalent to the hardest structural ceramics[4]. These abrasion- and load-resistant properties are due to the biogenic magnetite capping the tooth cusp[8]. In chitons, a single radula contains rows of teeth, with each tooth row progressing in discrete stages from an unmineralised to a fully mineralised state[9].

The major iron minerals that compose the structure of mature teeth in the Australian chiton *Acanthopleura hirtosa*[10] are magnetite, goethite, and calcium apatite (Fig. 1 b). Mineralisation is a controlled process, guided by an organic matrix[11,12], a framework composed primarily of the polysaccharide α-chitin[13]. Iron ions are complexed by the organic matrix[14], with ferrihydrite precipitating first, prior to its subsequent transformation to magnetite[15,16]. While the mechanical properties and mineral composition of mature chiton teeth have been studied extensively in multiple species[4,17–19], little is known about the growth and transformation of the iron biomineral phases and their organisation at the early stages of mineralisation. Fundamental questions remain on the mechanisms controlling ferrihydrite attachment to the underlying organic matrix and the subsequent transition of this precursor phase to magnetite. Additionally, the mechanisms that govern crystallisation to produce the hierarchical structure that imparts such mechanical strength in chiton teeth remain undiscovered. Addressing these questions is critical if we are to understand how these animals assemble such exquisitely fine-tuned materials.

To investigate the attachment, conversion and overall organisation of the iron biominerals at the early stages of development, we applied a recently developed quantum-based magnetic imaging approach[6,20,21] that combines high-resolution spatial imaging with quantitative mapping of static and fluctuating magnetic fields to study *in-situ* mineral phases. Our diamond-based magnetic microscope provides subcellular magnetic profiling while distinguishing the magnetic signals from magnetite and ferrihydrite.

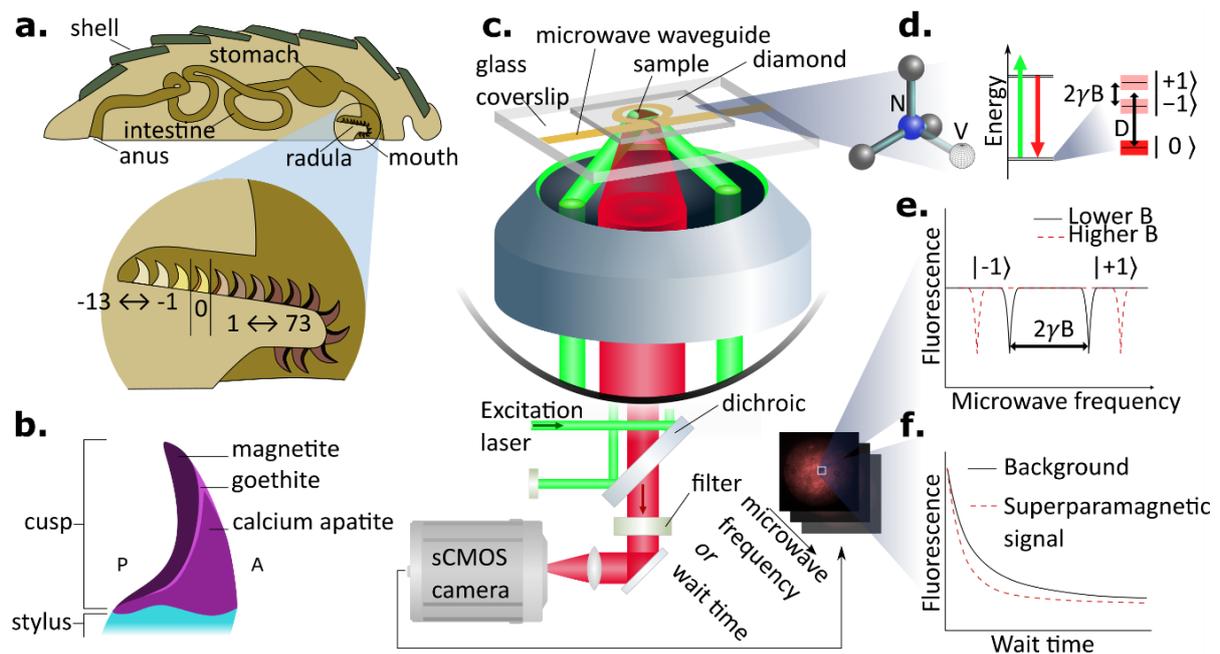

**Figure 1 | Quantum magnetic microscopy for studying iron-based biomineralisation**. **a**, Chiton sagittal section. Enlargement shows the radula – an organ hosting rows of teeth which develop in a conveyor belt-like fashion. The entire tooth development (87 tooth rows in *A. hirtosa*) is seen in one radula. **b**, The fully mature tooth cusp mineral structure of *A. hirtosa*, median longitudinal cut. **c**, Optical set-up of quantum magnetic microscope. The diamond sensing chip is excited via total internal reflection and the resulting fluorescence image is captured on an sCMOS camera. **d**, Schematic of the crystalline structure of the nitrogen vacancy (NV) centre in diamond, and relevant energy levels. **e**, Optically detected magnetic resonance schematic for imaging ferrimagnetic (magnetite) minerals. **f**, Quantum relaxation microscopy schematic for imaging superparamagnetic (ferrihydrite) minerals.

**Results**

To image the spatial distribution of iron-containing minerals, we designed and constructed a quantum-based magnetic microscope (QMM), which exploits the magnetic sensitivity of atomic defects in diamond[22]. Our system consists of an inverted optical wide-field microscope with a diamond sensing chip containing a 2D array of negatively charged nitrogen-vacancy (NV) centres[22] engineered approximately 6-8 nm beneath the diamond surface (Fig. 1c), see methods. A simplified schematic of the electronic structure of the NV centre is shown in Fig. 1d. The energy separation of the NV $|\pm1\rangle$ ground spin states is dependent on the local magnetic field $\Delta E = 2\gamma B_{NV}$, where $\gamma$ is the NV gyromagnetic ratio = 2.8MHz/G. The NV spins are optically polarised using green (532 nm) laser light, with the spin-dependent NV fluorescence (637-800 nm) imaged onto a sCMOS camera, allowing optically detected magnetic resonance (ODMR) to be performed at each imaging pixel of the sensing array[23]. We use two quantum sensing protocols to image the magnetic properties of chiton teeth at and before the onset of mineralisation: ODMR is used to image the static vector magnetic fields

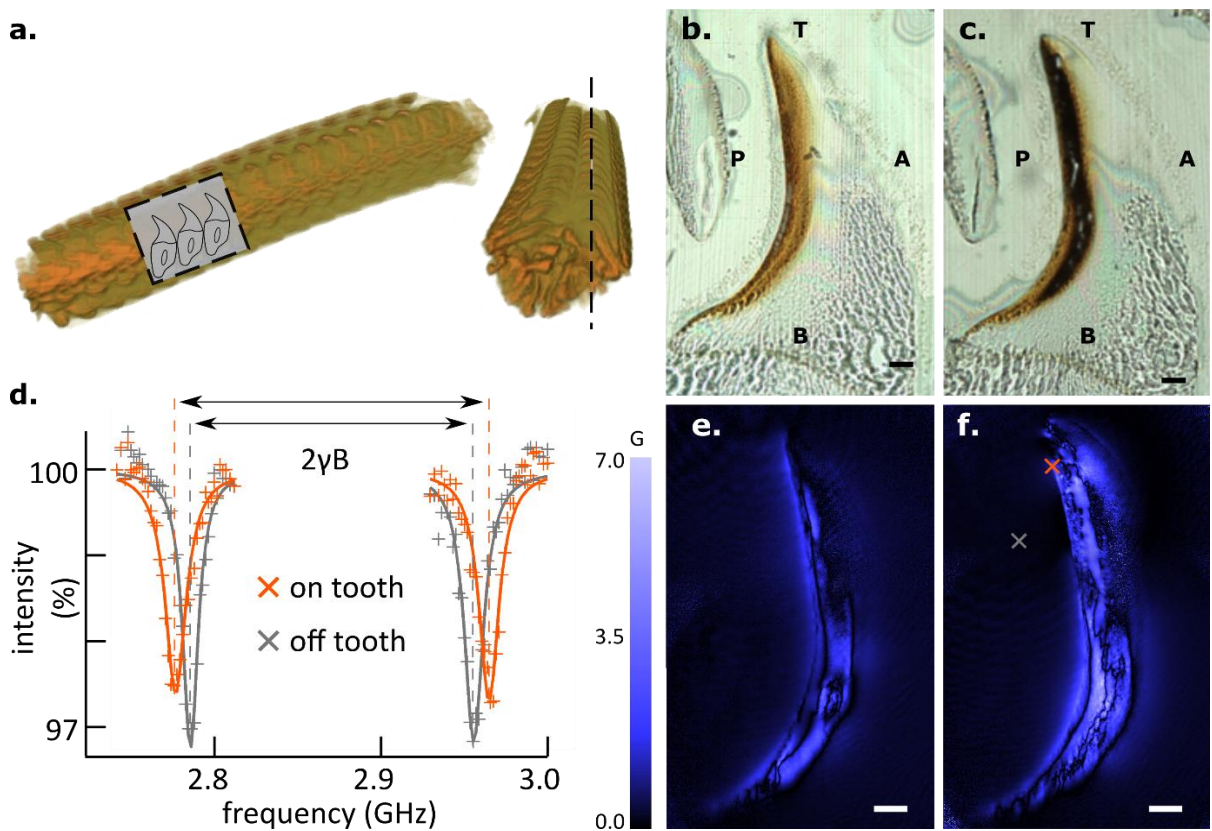

**Figure 2| Sample sectioning and magnetic imaging**. **a**, Radula set in resin, and microCT to facilitate longitudinal sectioning, dotted line. One micron thick slices of **b**, tooth 3, **c**, tooth 4, were mounted on the diamond sensing chip, scale bar 20 μm. **d**, Optically detected magnetic resonance (ODMR) spectra were measured at each imaging pixel across the entire tooth section. The separation of the ODMR transitions is proportional to local magnetic field $\Delta f = 2\gamma B$, where the gyromagnetic ratio of the NV centre $\gamma_0$ = 2.8 MHz/G. **e** and **f**, show the magnitude of

the measured magnetic field projection along a single NV axis at 109.5° from the normal for tooth 3 and 4, respectively. Crosses in **f** correspond to ODMR spectra in **d**.

from magnetite (Fig. 1e), and quantum relaxation microscopy to image the magnetic fluctuations from superparamagnetic ferrihydrite (Fig. 1f). Imaging may be performed consecutively without any adjustment of the sample or microscope setup.

*Acanthopleura hirtosa* radulae were set in resin and sectioned for imaging as shown in Fig. 2a. Correctly orienting the section through the tooth is crucial to achieve an accurate interpretation of the tooth structure. Because the radula provides a series of progressive snapshots of tooth development, we sectioned multiple teeth from the same radula. Teeth are numbered from less mineralised to more mineralised, with the first tooth that displays orange colouration being tooth zero. Sections were cut to show the tooth preceding mineralisation (tooth -1), at the onset of mineralisation (tooth 0), and more advanced stages of mineralisation where phase transformations to magnetite have occurred (teeth 3 and 4) (Fig. 2b and c).

The microtomed tooth sections were mounted on the diamonds for imaging, see methods. At room temperature, magnetite particles are ferrimagnetic and therefore present a static magnetic field that can be imaged via ODMR (Fig. 2d). The magnetic field strength from the magnetite nanoparticles can be measured via the splitting of the ODMR spectrum at each pixel to produce a 2D magnetic field map. All teeth were imaged in this mode, see supplementary information. No static magnetic signals were detected from tooth -1, indicating no ferrimagnetic magnetite nanoparticles, consistent with this stage of tooth development in *A. hirtosa*[11]. Magnetic signals corresponding to the posterior surface of tooth 0 were resolved, see supplementary information, and the strength of the static magnetic signal increases for both tooth 3 and 4 as shown in Fig. 2e and f. These magnetic images represent the magnitude of the magnetic field projection along a given NV axis. To determine the direction of the magnetite's magnetisation, we measure the full magnetic field vector. To implement this capability, we designed a three-axis Helmholtz coil to generate a weak and uniform background field of 30 G sequentially along each NV axis, allowing measurement of the magnetic field projection along each of the four NV crystallographic axes. We used the magnetic field projections along three NV axes to reconstruct the x, y and z projections (Fig. 3a and b) of the vector field.

The vector magnetic maps show a broadly uniform magnetite magnetisation, oriented perpendicular to the plane of the section (Fig. 3c). This long-range magnetic order is surprising, because at this stage of mineralisation, the magnetite particles are small and disconnected, without any orientational organisation evident in their individual visible structure alone, see supplementary information. Therefore, the long-range magnetic ordering indicates either that crystals were actively aligned by the organic matrix, or that the particles were magnetically coupled as they transitioned into a ferrimagnetic state. Disturbances in the magnetic field due to fractures in the tooth from sectioning are also clearly resolvable (Fig. 3d and e). A simplified magnetic model of tooth architecture supports the out of plane magnetisation finding, see supplementary information. Full vector maps of tooth 4 are also provided in the supplementary information with both teeth exhibiting similar magnetic profiles.

Previous work has identified anisotropy of magnetisation in the mature tricuspid teeth of *Acanthochiton rubrolinestus*,[24,25] where the length and width of the teeth are more easily magnetised than the thickness[26]. Our vector magnetic microscopy results demonstrate that for *A. hirtosa,* in the early stages of tooth development, the magnetite nanoparticles are magnetised perpendicular to the plane of the section, i.e. across the tooth width.

Biomineralisation has, in multiple instances, been shown to occur via non-classical routes, including mesocrystalline intermediate stages[27]. The possibility of a mesocrystalline formation mechanism was first raised for chiton biomineralisation upon the discovery of grain substructure in mature *A. hirtosa* teeth[7]. Our results support the presence of a magnetic mesocrystalline phase in the tooth cusp, preceding the ordered rods of magnetite in a mature tooth[16], as overall order is seen in the magnetic vector maps on a scale larger than any individual faceted crystal. It has been proposed that the collective magnetic field produced by the cumulative fields of aligned nanocrystals in a mesocrystalline colloid could assist the formation of a magnetic solid[28]. It is conceivable, therefore, that the magnetic field produced by the alignment of the early magnetite nanoparticles may assist in the synthesis of the magnetite rods.

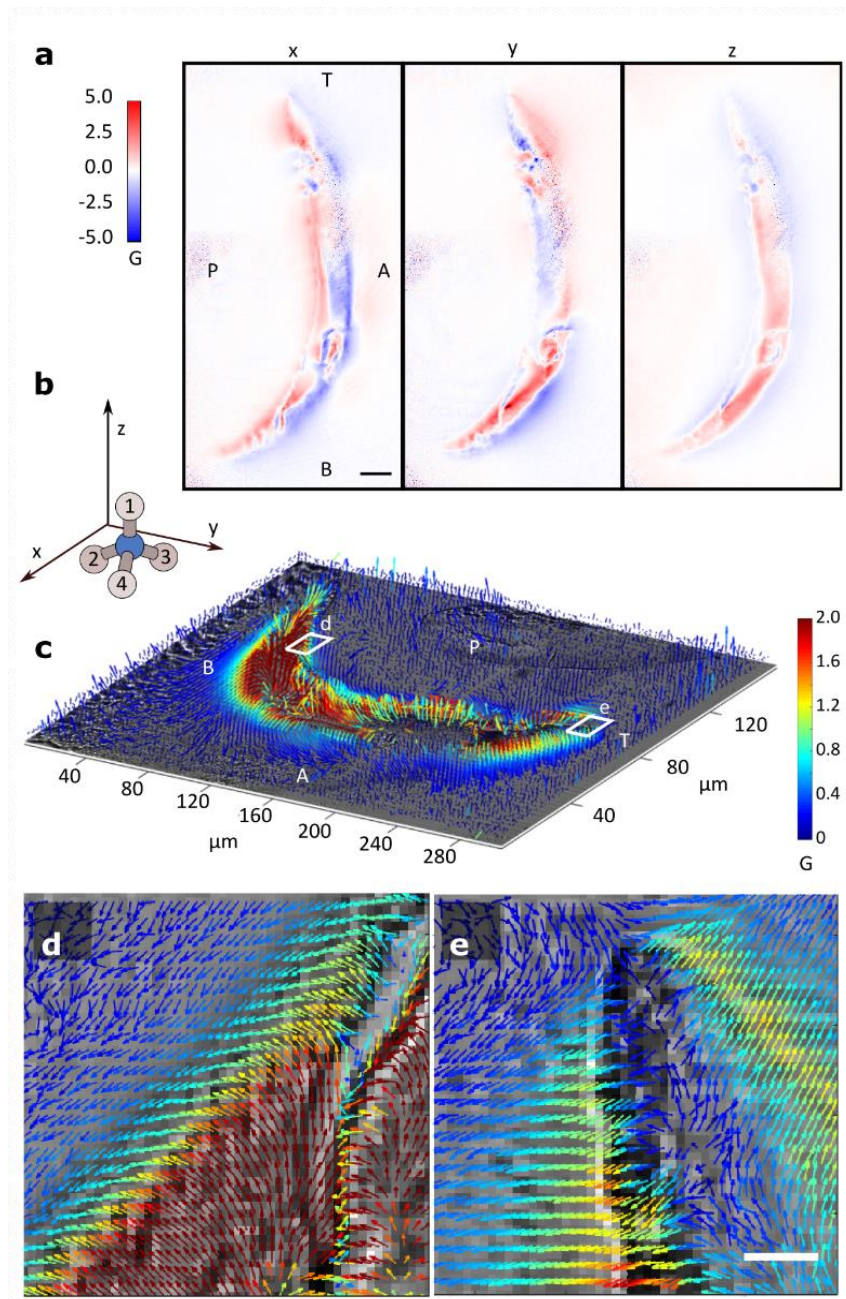

**Figure 3| Vector magnetic field imaging of chiton teeth**. **a:** maps of the x, y and z, (out-of-plane) components of the magnetic field, reconstructed from the four NV axes' maps. Letters indicate A anterior, P posterior, T tip, B base. **b:** Relative orientations of the diamond crystal axes and cartesian axes, as in figure c. **c:** Magnetic vector reconstruction of whole tooth cusp section of tooth 3. Colour scale represents magnitude of the magnetic field and the arrow direction represents magnetic field vector. To aid in visibility of signal from noise, a filter was applied reducing arrow length of noisy vectors. The vector reconstruction is performed by measuring the magnetic field projection along three known crystallographic directions of NV centres in diamond. The vector

reconstruction is performed at each imaging pixel with a spatial resolution dictated by the diffraction limit of the microscope, 300 nm. **d, e:** High resolution vector magnetic field maps of the x and y field components reveal long-range ordering along the tooth cusp, from base to tip, scale bar 2μm.

Prior to magnetite crystallisation, iron is present but not in a ferrimagnetic form. The first iron mineral to be deposited within the tooth is 6-line ferrihydrite, the predominant form of iron within ferritin[29,30]. Ferrihydrite is a ferric oxyhydroxide and superparamagnetic at room temperature[31]. Ferrihydrite particles are less than 10 nm in size[32], with a superparamagnetic fluctuation spectrum of order 1 GHz, see supplementary information. This magnetic fluctuation frequency conveniently overlaps with the NV transition energies from the $|0\rangle$ to $|\pm1\rangle$ states at zero magnetic field. By monitoring the spin relaxation rate ($1/T_1$) of the NV centres across the full field of view[33,34], see methods, we can detect and map the fluctuating magnetic fields from ferrihydrite directly, in a technique we term quantum relaxation microscopy (QRM).

Using QRM, we find the ferrihydrite concentration increases with development (Fig. 4a), in agreement with studies across chiton species[14,16]. In teeth -1 and 0, where little or no magnetite is present, ferrihydrite is detected along the posterior face of the tooth cusp. In the more mature teeth (3 and 4), the bulk of ferrihydrite is on the anterior side of the magnetite, with a thin layer of ferrihydrite coating the outer, posterior side of the magnetite (Fig. 4b and c).

As a mineral phase mapping technique, QRM is highly effective. Elemental mapping across whole *A. hirtosa* teeth has not previously been performed with any other technique; however, elemental and mineral phase locations have been investigated in other species. Energy dispersive spectroscopy and Raman spectroscopy have been used in combination with secondary electron microscopy to locate ferrihydrite and magnetite regions in the immature unicuspid teeth of *Acanthopleura echinata*[35]. Though not an imaging technique, the *A. echinata* results indicate multiple fronts of mineralisation, showing parallels with the present study. The QRM images show that by tooth 4, two distinct regions of ferrihydrite border the bulk of the magnetite region. In addition to the ferrihydrite band that appears on the posterior tooth cusp by tooth 4, the interior band extends further in towards the core of the tooth cusp. This contrasts with the ferrihydrite pattern in the tricuspid teeth of *Cryptochiton stelleri* observed via micro X-ray fluorescence (µXRF)[16]. These similarities and differences may be related to the mineralisation strategy employed; *A. echinata* and *A. hirtosa* form an apatite core while *C. stelleri* has a hydrated iron phosphate core. Because of the variation in these different mineralisation strategies, it would be worthwhile in future studies to examine multiple species with the high resolution, whole tooth mapping that QRM provides. The iron biominerals within chiton teeth studied here present two distinct magnetic signatures; however, more advanced stages of mineralisation will contain new biomineral phases such as goethite and lepidocrocite. Diamond-based electron spin resonance spectroscopy techniques[36] may allow these specific mineral phases to be mapped and imaged in more mature teeth.

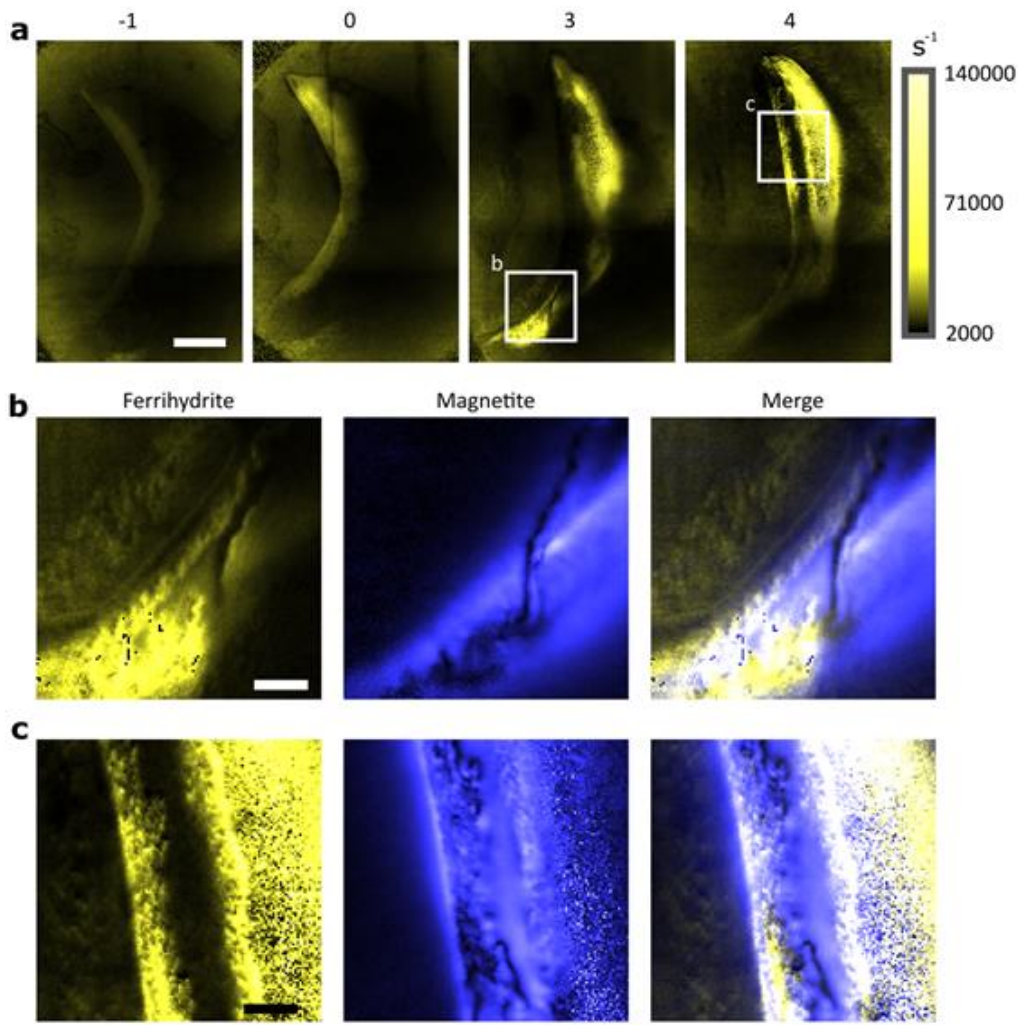

**Figure 4|** Superparamagnetic microscopy of chiton teeth. **a**, Quantum relaxometry images ($1/T_1$) of the NV centre in diamond showing the regions of superparamagnetic ferrihydrite. As the tooth develops, the total concentration and spatial distribution of ferrihydrite increases. Scale bar: 40 μm. **b** and **c**, show high resolution magnetic maps of ferrihydrite ($1/T_1$) - yellow and magnetite (ODMR) - blue in various sections of the tooth. Areas of co-localisation (white) reveal regions undergoing magnetic phase transformation. Scale bar: 10 μm

The application of quantum-based magnetic microscopy to biological systems presents several key advantages when compared to existing magnetic imaging techniques. Widefield magnetic microscopy allows for rapid acquisition and high throughput when compared to scanning probe techniques such as MFM and transmission electron microscopy. The unique combination of vector magnetic imaging and QRM on a bio-compatible platform opens a pathway toward mineral specific imaging of a vast range of biological systems. Additionally, these techniques could be adapted for real time monitoring of the assembly of engineered nanoscale magnetic materials.

**Methods**

**Materials**

The diamond imaging sensor used in this work is engineered from electronic grade Type IIa <111> diamond (Element 6). The diamonds were thinned, cut and re-polished to a 1 × 2 × 0.1 mm$^3$ crystal (DDK, USA). NV defects were engineered via ion implantation of $^{15}$N atoms at an energy of 4 keV and dose of 1 × 10$^{13}$ ions/cm$^2$. Molecular dynamic simulations indicate a NV depth range between 5-10 nm[37] . The implanted sample was annealed at 1000 °C for three hours and acid treated to remove any unwanted surface contamination. The density of NV centres post annealing was 1 × 10$^{11}$ NV/cm$^2$.

**Sample preparation**

Fresh specimens of the chiton *A. hirtosa* (Blainville, 1825) were collected and dissected as described previously[38]. Care was taken to avoid exposing the samples to any strong magnetic field from collection to measurement. Briefly, radulae were excised and fixed in 2.5% glutaraldehyde, buffered in 0.1 M phosphate at pH 7.2 (osmotic pressure adjusted to 900 mmol·kg−1 using sucrose). Radulae were then fixed, dehydrated, and infiltrated in epoxy resin using microwave-assisted chemical fixation (Pelco, Biowave).

To facilitate the collection of histological sections from precise orientations through the teeth, resin blocks were first scanned using X-ray micro-computed tomography (μCT). Scans were conducted at 80kV and 82 μA using a Versa 520 XRM (Zeiss, Pleasanton, USA) running Scout and Scan software (v11.1.5707.17179, Zeiss). A total of 401 projections were collected over 360º, each with a 1 second exposure. 2x binning was used to achieve a suitable signal to noise ratio and 0.4x optical magnification was used to achieve an isotropic voxel resolution of 12.8 μm. Raw data were reconstructed using XMReconstructor software (v11.1.5707.17179, Zeiss) following a standard centre shift and beam hardening correction. The standard 0.7 kernel size recon filter setting was also used. Reconstructed data were observed (TXM3D Viewer, Zeiss) to determine the position of the major lateral teeth relative to the resin block and a razor blade was used to then mark the resin at the correct plane of orientation to produce a longitudinal cut through the middle of each tooth.

Resin blocks were then trimmed at this orientation using a glass knife on an ultramicrotome (EM UC6, Leica Microsystems). Once the correct position was reached, 1 µm sections were cut with a diamond knife (Histo, Diatome) onto filtered DI water. Sections were transferred using a wooden applicator stick to a drop of ~70 °C filtered DI water (situated in a Petri dish placed on a hotplate). Sections were left to warm for 1-2 mins on the water drop to assist with smoothing the section. Sections were transferred using a wooden applicator stick to a drop of filtered DI water on the imaging slide. An eyelash mounted on a wooden applicator stick was used to position the section onto the diamond chip and a paper wick was then used to remove excess water from the slide; finally, the sample was air dried.

**Optical Imaging**

Samples were diamond-imaged in wide-field on a modified Nikon inverted microscope (Ti-U). Optical excitation from a 532 nm Verdi laser was focused (f = 300 mm) onto an acousto-optic modulator (Crystal Technologies Model 3520–220) and then expanded and collimated (Thorlabs beam expander GBE05-A) to a beam diameter of 10 mm. The collimated beam was focused using a wide-field lens (f = 300 mm) to the back aperture of the Nikon x60 (1.4 NA) oil immersion objective via a Semrock dichroic mirror (Di02-R561-25 × 36). The beam was first centred to the objective, then translated until the beam totally internally reflected within the diamond. The NV fluorescence was filtered using two bandpass filters before being imaged using a tube lens (f = 300 mm) onto a sCMOS camera (Neo, Andor). Microwave excitation to drive the NV spin probes was applied via an omega gold resonator (diameter=0.8mm) lithographically patterned onto a glass coverslip directly under the diamond imaging chip. The microwave signal from an Agilent microwave generator (N5182A) was switched using a Minicircuits RF switch (ZASWA-2-50DR+). The microwaves were amplified (Amplifier Research 20S1G4) before being sent to the microwave resonator. A Spincore Pulseblaster (ESR-PRO 500 MHz) was used to control the timing sequences of the excitation laser, microwaves and sCMOS camera and the images where obtained and analysed using custom LabVIEW code. The excitation power density used for imaging was 30 W/mm$^2$ and all images were taken in an ambient environment at room temperature.

**Quantum relaxation microscopy**

To implement quantum relaxation microscopy the spin lattice relaxation time ($T_1$) of the NV centres was determined by optically polarising the NV spins into the $m_s = 0$ ground state, then allowing the spins to evolve (in the dark) for a time $\tau$, before sampling their spin polarisation with an additional optical pulse. Interactions between the NV centres and neighbouring electronic, nuclear and surface spins species cause the NV net magnetization to relax from the $m_s = 0$ state to a mixture of the three ground triplet states. The $e^{-1}$ time of the decay is the $T_1$ time of the NV centres. In this work we normalised the $T_1$ decay with an identical pulse sequence with a single $\pi$ pulse applied prior to the spin readout. This provides common mode rejection of noise sources from the NV imaging array. The quantum relaxation microscopy image analysis was performed using custom LabVIEW code.

**Electromagnet**

Magnetic fields were applied with a custom 3D Helmholtz coil to control both the strength and the direction of the magnetic fields. Copper enamelled wires (ECW0.71) were wound 200-time rounds on each cylindrical side of the Helmholtz coil. To reduce temperature increasing caused by the Joule heating of the copper wires, the coil was manufactured from aluminium and a computer water cooling system (EK CoolStream PE 240 Dual Radiator, EK Vardar 120mm Fan F4-120 2200RPM, and EK-XRES 100 SPC-60 MX PWM Pump/Reservoir Combo) was installed, with water running through a path inside the coil. The power supplier (Hameg HMP2030) was used for generating a static magnetic field.

**Image analysis**

Custom LabVIEW code was written for image analysis. To produce static magnetic field images, NV fluorescence images were binned 4x4, and ODMR spectra across the two peaks corresponding to the aligned NV axis were obtained for each binned pixel of the image stack. The spectra were fitted with single Lorentzian and the magnetic field strength determined from the peak splitting. Magnetic field strength at each binned pixel was saved as an image, and a colour scale applied with ImageJ (Fiji distribution, ImageJ 1.51h). To produce the quantum relaxation images, the NV fluorescence images were binned 4x4, and the $T_1$ decay curves were obtained for each binned pixel of the image stack. The $T_1$ decay rate ($1/T_1$) was determined at each binned pixel by fitting the data to a stretched exponential of the form $y = A \, exp^{(t/T_1)^p} + c$, where A is the amplitude of the exponential decay, $T_1$ is the spin lattice relaxation time, p is the stretched exponential power (p = 1 represents a single exponential decay) and c is the offset. Near surface NV centres are known to exhibit a distribution of $T_1$ times from the NV ensemble depending on their proximity to the surface and local spin environment. This distribution leads to a non-exponential $T_1$ decay which is characterised well by a stretched exponential function. The $T_1$ rate at each pixel binned pixel was saved as an image, and a colour scale applied with ImageJ (Fiji distribution, ImageJ 1.51h).

**Acknowledgements**

This work was supported by the Australian Research Council (ARC) through Grants No. FL130100119, DE170100129. J.-P.T. and L. T. Hall acknowledges support from the University of Melbourne through an Early Career Researcher Grants. The authors acknowledge the facilities, and the scientific and technical assistance of Microscopy Australia at the Centre for Microscopy, Characterisation & Analysis, The University of Western Australia, a facility funded by the University, State and Commonwealth Governments.

**Author Contributions**

D.A.S., J.M.M., J.S. and L.C.L.H. conceived the study. D.A.S. designed and constructed the quantum magnetic microscope. J.S. collected the chiton samples for the study, performed μCT imaging and sectioned and mounted the samples onto the diamond imaging chips. J-P.T. and D.S. prepared the diamond imaging chips. J.M.M. performed the ODMR and $T_1$ imaging with assistance from M.M. and R.W.deG. M.M. designed and fabricated the Helmholtz coil for the vector magnetic imaging. D. K. provided key insights into the mechanisms driving biomineralisation in chiton teeth. The theoretical framework for the imaging was established by L.T.H and L. C. L. H. J.M.M. developed the magnetic model used in the magnetic simulations with assistance from J-P.T. D.A.S. and L. C. L. H. supervised the project. All authors discussed the results and participated in writing the manuscript.